
\input harvmac

\Title {DAMTP93/R20}
{{\vbox {\centerline{The Quantum Stress Tensor} \break
 \centerline{in the Three Dimensional
Black Hole }}}}

\bigskip
\centerline{Alan R. Steif}
 \bigskip\centerline{\it Department of Applied Mathematics and Theoretical
Physics}
\centerline {\it Cambridge University}
\centerline{\it Silver St. }
 \centerline{\it Cambridge, U.K. CB3 9EW }
\centerline{\it ars1001@moria.amtp.cam.ac.uk}
\vskip .2in

\def\Ads3{ADS$_3$}
\def\L{\Lambda}
\def\A{\alpha_+}
\def\B{\alpha_-}
\def\b{\beta}
\def\p{\phi}
\def\na{\nabla}

\def\d{\delta}
\def\gmn{g_{\m\n}}

\def\l{\lambda}
\def\xpr{x^{\prime}}
\def\qem{ <{\rm T}_{\m\n}>}
\def\em{ T_{\m\n}}
\def\ra{\rightarrow}
\def\a{\alpha}
\def\b{\beta}
\def\pa{\partial}
\def\tt{\tilde{ t}}
\def\tr{\tilde {r}}
\def\tp{\tilde {\phi}}

\def\Nm{\nabla_{\mu}}
\def\NM{\nabla^{\mu}}

\def\xpr{x^{\prime}}

\def\cp{\cosh\,n\A}
\def\cm{\cosh\,n\B}
\def\r{\rho}

 \def\n{\nu}
\def\m{\mu}

\noindent
ABSTRACT:
The quantum stress tensor $<T_{\mu\nu}>$ is calculated   in the 2+1 dimensional
black hole found
by Banados, Teitelboim, and Zanelli. The Greens function,
 from which  $<T_{\mu\nu}> $   is derived,   is  obtained by  the method of
images.
 For the  non-rotating black hole, it is shown that
 $<T_{\mu\nu}>$    is finite on   the event horizon, but diverges at the
singularity.
For the rotating solution, the stress tensor   is finite at the outer horizon,
but diverges near the inner horizon. This suggests that the inner horizon is
 quantum mechanically unstable against the  formation of a singularity.
\Date{}
 \eject

Recently, Banados,  Teitelboim, and  Zanelli {\ref\btz{
M. Banados, C. Teitelboim, and J. Zanelli, {\it Phys. Rev. Lett.} {\bf 69}
(1992) 1849;  M. Banados, M. Henneaux,  C. Teitelboim, and J. Zanelli,
 {\it Phys. Rev. D} {\bf 48} (1993) 1506.}}
found a black hole solution in  $2+1$ dimensions
 which  shares many of the features of its $3+1$ dimensional counterpart
{\ref\others{S. F. Ross and R. B. Mann, {\it Phys. Rev. D} {\bf 47} (1993)
3319;
 G.T. Horowitz and D.L. Welch,  {\it Phys. Rev. Lett.} {\bf 71} (1993) 328;
 N. Kaloper,  {\it Phys. Rev. D} {\bf 48} (1993) 2598;
A. Achucarro and M. Ortiz,  {\it Phys. Rev. D} {\bf 48} (1993) 3600;
  D. Cangemi, M. Leblanc, and R. B. Mann, {\it Phys. Rev. D} {\bf 48} (1993)
3606;
C. Farina, J. Gamboa, and A.J. Segui-Santonja, preprint IPNO-TH 93/6.
 }}.  In particular, the static solution  has a singularity and event horizon,
while the rotating black hole
like  Kerr  possesses outer and inner horizons and an   ergosphere.
Asymptotically, however, the $2+1$ solution  is not flat,
but approaches anti-deSitter space {\ref\hp{Black holes in $3+1$ anti-deSitter
 space have been discussed in S. W. Hawking and D. Page,
{\it Comm. Math. Phys.} {\bf 87} (1983) 577.}}.
 $2+1$ dimensions   provides a simpler setting than $3+1$  and possibly a more
 realistic one than $1+1 $
{\ref\hs{For a review, see J. Harvey and A. Strominger,
``Quantum Aspects of Black Holes'', lectures given
 at  the 1992 TASI Summer School.}}
 in which to study the quantum properties of black holes,
and specifically,   the endpoint of black hole evaporation.
  Such an investigation should begin with the quantum  stress tensor
$<T_{\m\n}>$  which describes the quantum effects of the black hole on a
propagating field
in a way that allows one to   analyze the back reaction.
 Provided it can be properly renormalized, $<T_{\m\n}>$  is a well defined
local quantity in contrast to  particle number which  is not, in general,
 a meaningful concept in curved spacetime.
Another motivation for studying $<T_{\m\n}>$  in the rotating black hole
  is to investigate the quantum stability of the inner horizon. The maximally
extended Reissner-Nordstrom and Kerr solutions include an infinite number of
 asymptotic regions which in principle could be accessed. However,
 it has been shown that since the inner horizon is
an infinite blueshift surface, classical perturbations will diverge there
 {\ref\clin{M. Simpson and R. Penrose, {\it Int. J. Theor. Phys.} {\bf 7}
(1973) 183;  J. McNamara,
 {\it Proc. Roy. Soc. Lond.} {\bf A358} (1978) 499; R. Matzner, N. Zamorano,
 and V. Sandberg, {\it Phys. Rev. D} {\bf 19} (1979) 2821;
N. Zamorano,  {\it Phys. Rev. D}  {\bf 26} (1982) 2564;
S. Chandrasekhar and J. Hartle,
{\it Proc. Roy. Soc. Lond.} {\bf A384} (1982) 301.}},
 and the associated back reaction will produce a singularity
 {\ref\pis{  W. Hiscock, {\it Phys. Lett. A} {\bf 83} (1981) 110;
E. Poisson and W. Israel, {\it Phys. Rev. D} {\bf 41} (1990) 1796; A. Ori,
 {\it Phys. Rev. Lett.} {\bf 67} (1991) 789;  {\bf 68} (1991) 2117. }}.
Quantum effects for
the $1+1$ dimensional analog of the   Reissner-Nordstrom solution  were
investigated in  {\ref\his{
W. Hiscock, {\it Phys. Rev. D} {\bf 15} (1977) 3054; N. Birrell and P. Davies,
{\it Nature} {\bf 272} (1978) 35.}} where it was  shown that
 $<T_{\m\n}>$  diverges  near the inner horizon.
 Attempts to include quantum corrections in $3+1$ dimensions
  {\ref\bp{R.  Balbinot and E. Poisson, {\it Phys. Rev. Lett.} {\bf 70} (1993)
13;
 W. Anderson, P. Brady, W. Israel, and S. Morsink, {\it Phys. Rev. Lett.}
{\bf 70} (1993) 1041; W. Israel, in {\it Directions in General Relativity,}
 vol. I,  eds. B. Hu, M. Ryan Jr., and C. Vishveshwara, (Cambridge Press,
1993).}}
are somewhat inconclusive suggesting that  the classical instability  either is
enhanced
or is dampened resulting in a regular spacetime.
 In this paper, the exact expression for  the quantum stress tensor is found
for  the rotating $2+1$ dimensional black hole and is shown to diverge
near the inner horizon. An estimation of the back reaction suggests
that the inner horizon will be replaced by a curvature singularity.  We use
units in which
$\hbar = c= G =1$.

 The $2+1$ dimensional black hole solution  found by  Banados,
 Teitelboim, and  Zanelli  {\btz} is most easily described as
  three dimensional anti-deSitter space  (\Ads3)  identified
under a discrete subgroup  of its  isometry group. Recall that {\Ads3} is the
three dimensional hypersurface
\eqn\ads{
-T_1^2 + X_1^2 - T_2^2 +  X_2^2 = -l^2
}
imbedded in   four dimensional flat space with metric $\eta_{ab}$
\eqn\imbmet{
ds^2 = - d T_1^2 + d X_1^2 - d T_2^2 + d X_2^2
 }
where $l = (- \Lambda)^{-1/2}$.
The hypersurface {\ads} is a pseudohyperbolic analog of a three sphere
with radius vector $x^{a } \equiv  ( T_1, X_1, T_2, X_2) ,$   radius $ \sqrt{ -
x^ax_a} =  l ,$
and  constant curvature $R= -6/l^2$.
We will use lowercase Latin indices for the four dimensional
imbedding space and lowercase Greek indices for {\Ads3}.
The isometry group of {\Ads3} is $ SO(2,2)$
and corresponds to the subgroup of the isometry group of the
 imbedding space which leaves {\ads} invariant. Since boosts
and rotations in two dimensional planes generate the isometry group,
the simplest coordinate systems for {\Ads3}    parameterize these
symmetries. As we will see, the black hole solution
is constructed by identifying the parameters describing boosts
in the  $(T_1, X_1)$ and   $(T_2, X_2)  $ planes. Thus,
it is in terms of these boost parameters that we wish to express the metric for
{\Ads3}.
 We view it in terms of two copies of $1+1$ Minkowski space,
 $M_1$ with coordinates $(T_1, X_1)$ and $M_2$  with coordinates $(T_2, X_2)
,$
 with the constraint {\ads}   $\r_1 + \r_2 = l^2 $ where
 $\r_i = T_i^2 - X_i^2$. In each space $M_i$, one can define Rindler
coordinates
\eqn\trphicoor{\eqalign{
T_i &=  \sqrt{\r_i} \cosh \, \chi_i,  \quad\quad \;\; X_i  =
 \sqrt{\r_i} \sinh \, \chi_i, \quad \;\; \r_i >0, \;\; -\infty < \chi_i <
\infty \cr
 T_i &=  \sqrt{-\r_i} \sinh \, \chi_i,  \quad\quad  X_i =
 \sqrt{-\r_i} \cosh \, \chi_i, \quad \r_i <0, \;\; -\infty < \chi_i < \infty
\cr
}}
valid in the lightcone interior $(\r_i >0 )$ and  exterior  $ (\r_i <0 )$
respectively.
 Defining $\chi_1 \equiv \phi$ and $\chi_2 \equiv t$, we see that
  there are three qualitatively distinct regions:
$(I)$ \quad  $\r_1 > l^2  $  $(\r_2 < 0),$  \quad $(II)$ \quad $0< \r_1
, \r_2 <l^2 ,$ and $\;\;(III)$  \quad $\r_1  <0  $ $(\r_2 > l^2)  ,$
in which the  vectors ${\pa\,\over \pa\p}$ and  ${\pa\,\over \pa t}$ are
 spacelike and timelike, spacelike and spacelike, and timelike and spacelike,
respectively.
It is   natural to view $I$ as the asymptotic region of the spacetime.
 Substituting {\trphicoor}  in {\imbmet} with $r^2 \equiv \r_1 = l^2 - \r_2 $,
 one obtains the metric for {\Ads3}
\eqn\trpmetric{
ds^2  = - ({r^2\over l^2} -1) dt^2 + ({r^2\over l^2} -1)^{-1} dr^2 + r^2 d \p^2
, \quad t, \p \in ( - \infty, \infty )
}
valid in regions $I$ and $II$. Since $t$ and $\p$
parameterize boosts, they take on all real values.

The black hole solution is now constructed by making some   combination of
$\p$ and $t$ periodic.
 For the static black hole with mass $M$, one  identifies
 $\p$ with period  $2\pi \sqrt{M}$.  This is somewhat analogous
 to the identification which leads to the static cone solution
in $2+1$ gravity without a cosmological constant
\ref\djt{S. Deser, R. Jackiw, and G.'t Hooft, {\it Ann. Phys.} (NY)
{\bf 152}  (1984) 220; S. Deser and R. Jackiw, {\it Ann. Phys.} (NY) {\bf
153} (1984) 405.}. A salient difference, however, is that  the cone
reduces to  flat space as $M\ra 0$, while {\Ads3},
the covering space of the black hole, is recovered as $M\ra \infty$.
One would expect the event horizon and singularity of the black hole
to have a natural geometric interpretation in terms of {\Ads3}.
Indeed, the event horizon  is located at   $(r=l)$    and  coincides with the
boundary between regions
$I$ and $II$ in {\Ads3} as well as with the light cone in the $1+1$
space $M_2$.  The black hole singularity is located at  $  r =0 $ corresponding
to  the boundary  between regions
$II$ and $III$ and to the light cone in $M_1$.   $r=0$ is not
 a curvature singularity   since the   curvature is bounded and in fact,
constant in {\Ads3}.  It is however   a singularity because
  there are inextendible incomplete   geodesics.   $r=0$ is directly analogous
to the   Misner space light cone \ref\he{S. W. Hawking and G. F. R. Ellis, {\it
The Large Scale Structure of Spacetime} (Cambridge University Press, 1973).}
on which  incomplete null geodesics  pile up.
Asymptotically, the black hole solution approaches anti-deSitter space.

The  black hole with non-zero angular momentum  $J$ is obtained
from {\trpmetric} by making  a linear combination of $\p$ and $t$ periodic:
$ (t,r,\phi)  \sim (  t  -   n l \B , r, \phi +  n \A ) $   where
\eqn\ab{
\a_{\pm} = {\pi } ( \sqrt {M+J/l} \pm \sqrt{M-J/l} ).
}
  It is possible to transform to  coordinates $(\tt, \tr, \tp) :$
 \eqn\trans{\eqalign{
t &=  {1\over 2\pi} ({\a_+} \tt - {\a_-}l \tp )\cr
\p &=  {1\over 2\pi} ( {\a_+ } \tp - {\a_-} \tt/l )\cr
r^2 &  = { (2\pi  \tr )^2 - {\a_-}^2  l^2\over {\a_+}^2 - {\a_-}^2 }\cr
}}
in terms of which   the metric  {\trpmetric} becomes
\eqn\ttrpmetric{
ds^2 =  - ({\tr^2 \over l^2}-M) d\tt^2 - J d\tt d\tp   + ({\tr^2\over l^2} -
 M + {J^2 \over 4\tr^2} )^{-1} d\tr^2 + \tr^2 d{\tp}^2
}
and $\tp $ is periodic in $2\pi$.
 The rotating solution possesses  both  an outer and inner horizon at
$\tilde r=  \a_+ l/ {2}\pi$ $(r=l)$ and $\tilde r= \a_- l/{2}\pi$ $(r=0)$
corresponding respectively to the   boundaries between regions $I$ and $II$
and between $ II $ from $III $ in {\Ads3}.  In addition,  the region
$  \a_+ l/ {2}\pi < \tr <  {\sqrt {M} } l$ defines    an ergosphere,
 in which   the asymptotic Killing field ${\pa\,\over \pa \tt}$ is spacelike.
 Finally, one should note that in contrast to the static $J=0$ black hole,
 the rotating solution is geodesically complete.

The points identified in the rotating    black hole are related by an
element of    $ SO(2,2)$ which as a matrix acting on the imbedding space
coordinates
 $x^{a } \equiv  ( T_1, X_1, T_2, X_2) $ takes the form
 \eqn\lorentz {
 \L^{a}_{\,\,b}  \equiv {\pmatrix{ \cosh\, \a_+ & \sinh \, \a_+ & 0 &0\cr
                               \sinh \, \a_+     &  \cosh\, \a_+ &0&0\cr
                               0&0&   \cosh\, \a_- &  -\sinh \, \a_-  \cr
                               0&0&   -\sinh \, \a_- &  \cosh\, \a_-\cr }
}
.}
For $J=0$ $( \a_- =0 )$, $\L$ reduces to a boost in the $M_1$ space, or
equivalently a translation in $\p ,$ and has   fixed points coinciding with
 the singular surface  $r=0$. For $J\neq 0$, $\L$ has
no fixed points accounting for the non-singular nature of the rotating
solution.

We now introduce a propagating quantum field in the black hole background and
 calculate  its Greens function.  Consider a conformally coupled massless
scalar field $\p$ governed by the action
   \eqn\action{
S = - \int  ( {1\over 2}(\nabla \p)^2  + {1\over 16}  R \p^2) \sqrt {g} d^3x
}
with  $R$   the scalar curvature.  We first review the construction of
the Greens function in {\Ads3},
 the covering space of the black hole {\ref\ais{S. Avis, C. Isham, and
D. Storey, {\it Phys. Rev. D} {\bf 18} (1978) 3565.}}. {\Ads3} is a static
spacetime with a globally defined timelike Killing field corresponding to
the generator of rotations in the $(T_1, T_2)$ plane in the imbedding
space. There is therefore a natural  vacuum state defined by modes
which are positive frequency with respect to this time parameter.
Since anti-deSitter space is not globally hyperbolic, it is
important to address the issue of boundary conditions at infinity.
 {\Ads3}  can be conformally
mapped to half of the Einstein static universe with infinity mapped to the
equator {\he}. Therefore, solutions to the equations of motion
in one space can be mapped to solutions in the other, and similarly, boundary
conditions at infinity correspond
to conditions on the fields at the equator.
As discussed in {\ais}, there are three natural choices of boundary conditions.
The first which is known as ``transparent'' simply corresponds to   quantizing
the field using modes which are smooth on the entire
Einstein static universe.
The other two boundary conditions are obtained by imposing
Dirichlet or Neumann conditions on the field
at the equator in the Einstein static universe.
 The Greens function is given by
\eqn\bargreengen{
\bar G_{\lambda}(x, \xpr) =  {1\over 4\pi} {1 \over | x - \xpr|} +  {\lambda
\over 4\pi} {1 \over | x + \xpr|}  }
with $\lambda = 0, 1, -1$ for transparent, Neumann, and Dirichlet boundary
conditions respectively.
 Observe that   $ | x - \xpr|
\equiv ( ( x - \xpr)^a  ( x - \xpr)_a )^{1/2}$ is the chordal distance
between $x$ and $\xpr$ in the four dimensional imbedding space and not the
distance in {\Ads3}. The second term in {\bargreengen} is obtained from the
first by the antipodal transformation
$\xpr \rightarrow - \xpr$, a discrete isometry of {\Ads3}.
In this paper, we will be considering only the $\lambda =0 $ Greens function
  corresponding to transparent boundary conditions
\eqn\bargreen{
\bar G(x, \xpr) =  {1\over 4\pi} {1 \over | x - \xpr|} .  }
Note that the Greens function    coincides with its form in   three-dimensional
Minkowski space.
 This is expected as  $\p$ is conformally coupled and  {\Ads3} is conformally
flat.
  We now verify that the Greens function     satisfies the $\p$ equation of
motion as derived from {\action}
   \eqn\greeneqn{
(\NM \Nm  +   {3/4\over l^2}) \bar G (x, \xpr) =  0,\quad x \neq \xpr .}
This is most easily checked by    expressing the wave operator $ \NM \Nm $
in  {\Ads3}   in terms of derivatives
$\pa_a$ in the imbedding  space.
 $P^{ab} = \eta^{ab} + x^a x^b / l^2$ satisfies $P^{ab} x_b =0 $ and is a
 projection operator for {\Ads3}. Applying it to the wave operator $ \pa^a\pa_a
$, one obtains
\eqn\waveop{\eqalign{
 \NM \Nm  & = P^{ab}    \pa_a ( P_b^{\, c} \pa_c)\cr
 & = P^{ab} \pa_a \pa_b  + 3{x^a \over l^2} \pa_a .\cr
}}
Using this one verifies that {\bargreen} satisfies {\greeneqn}.
 Since the black hole solution corresponds to {\Ads3} with discrete
identifications,
the Greens function $ G(x, \xpr)$ for  the black hole  can be obtained
from the Greens function {\bargreen} for  its covering space
by the method of images {\ref\sm{   In  K. Shiraishi and T. Maki,
``Vacuum Polarization around a Three-Dimensional Black Hole''
Akita Junior College preprint AJC-HEP-18,   the   Greens function for
 the  Euclidean continuation of  the non-rotating black hole  was
calculated using   a different technique.    Its Lorentzian continuation
agrees with the Greens function derived here.}}. Since the images
of $\xpr$ are $\L^n \xpr$ with $\L$ given in {\lorentz}, the Greens function
is
\eqn\green{
  G(x, \xpr) =  \sum_{n= -\infty}^{\infty}  \bar G(x, \L^n \xpr) =
{1\over 4\pi} \sum_{n= -\infty}^{\infty}    {1 \over | x - \L^n \xpr|} .}
  The contributions  from the $n$th and $-n$th terms insure that {\green} is
symmetric in $x$ and $\xpr .$

The quantum stress tensor can now be  obtained from $ G(x, \xpr)$.
Varying the action {\action} with respect to   $ g_{\m\n}$ yields
 \eqn\em{
T_{\m\n}  = {3\over 4}   \na_{\m}\p \na_{\n} \p - {1\over 4}\gmn (\nabla \p)^2
- {1\over 4}\p \na_{\m}
\na_{\n} \p + {1\over 4} \gmn  \p \na^{\l} \na_{\l} \p  + {1\over 8} G_{\m\n}
\p^2
}
with  $ G_{\m\n}$,  the Einstein tensor for the background spacetime.
 It follows from the  equation of motion  for  $\p$ that $T_{\m\n}$  is
traceless and conserved.
 The
quantum stress-tensor  $ <T_{\m\n}>$  is obtained  by  point splitting {\em}
and then
taking its  expectation value.   Using the $\p$ equation of motion in the
fourth term,
and substituting  in $ G_{\m\n} = l^{-2} \gmn$  for {\Ads3}, one obtains
\eqn\quem{
{\qem}  = {\rm lim}_{\xpr\ra x} ({3\over 4}   \na^x_{\m}  \na^{\xpr}_{\n} G
-  {1\over 4} \gmn  g^{\a\b} \na^x_{\a}  \na^{\xpr}_{\b} G
-   {1\over 4}\na^x_{\m}  \na^{x}_{\n} G  - {1 \over 16 l^2 } \gmn G
 )}
in terms of the    Greens function {\green}.   The renormalization of the
stress
 tensor,  ordinarily a difficult procedure  in  $3+1$ dimensions
{\ref\hc{K. Howard and P. Candelas, {\it Phys. Rev. Lett.} {\bf 53} (1984) 403;
K. Howard, {\it Phys. Rev. D} {\bf 30} (1984) 2532.}},  is achieved here by
simply
subtracting off the coincident $n=0$ term in the image sum {\green}
   \ref\hk{W. Hiscock and D. Konkowski, {\it Phys. Rev. D} {\bf 26}  (1982)
1225; J. Grant, {\it Phys. Rev. D} {\bf 47} (1993) 2388.
}.
Substituting {\green } into {\quem} and using $\na_{\m}  \na_{\n} x^a =
\gmn x^a / l^2 ,$
  one eventually
finds
\eqn\tmn{\eqalign{
<T_{\m\n}>&  = {3\over 16 \pi } \sum_{n \neq  0}
(  S^n_{\m\n} - {1\over 3} \gmn g^{\lambda\rho}
S^n_{\lambda\rho} ) \cr
 S^n_{\m\n} = \pa_{\m}x^a \pa_{\n}x^b  S^n_{ab}, \quad
 S^n_{ab} & =  {(\L^n)_{ab} \over  |x - \L^n x|^3}  +  {3 (\L^n)_{ac} x^{c}
(\L^{-n})_{bd} x^{d} - (\L^n)_{ac} x^{c}
(\L^n)_{bd} x^{d}\over  |x - \L^n x|^5 } . \cr
 }}
   $ S^n_{\m\n}$ is the pull back to {\Ads3} of  $ S^n_{ab}$.

The stress tensor {\tmn} can be evaluated in a particular set of coordinates
 $y^{\m}$ in {\Ads3} by substituting in the corresponding imbedding $x^{a }=
x^{a }( y^{\m}).$
  For the  static $J=0 \; (\a_- =0)$ black hole in coordinates $(t, r, \p )$
{\trpmetric}, {\tmn} takes the form
\eqn\stressjzero {
 {<T_{\m}^{\n}>}=  {A(M)\over {r}^3} {\rm diag} (1, {1}, {- 2}), \quad A(M)
\equiv
{\sqrt{2}\over 32\pi}   \sum_{n =1}^{\infty} { \cosh\, 2n\pi\sqrt{M} + 3\over
(\cosh\, 2n\pi\sqrt{M} - 1)^{3/2}}
}
  where $M$ is the black hole mass. Since the series  converges exponentially
for all real $M$,
the stress tensor is finite everywhere except near the singularity
where it diverges as $r^{-3}$. The divergence  there arises from the fact
that since $r=0$  remains invariant under the action of $\L$, the denominator
in the Greens function {\green} vanishes. Even though the coordinates $(t, r,
\p )$
 breakdown near the event horizon, it is clear that the scalar
$<T_{\m\n}><T^{\m\n}>$ is smooth there.
   For    $M >>1$,     the first term in the series
 gives  the leading order behavior $ A(M) \sim e^{-\pi \sqrt{M}}$.
Recall that as $M\ra \infty $, $\p$ becomes unidentified and {\Ads3} is
 recovered. Since $<T_{\m\n}>$ was renormalized with respect to  {\Ads3},
 it vanishes in this limit. For small $M$, the series can be approximated by an
integral yielding
$A(M) \sim  M^{-3/2}$.
{}From the invariance of the vacuum under the anti-deSitter group,
one would expect $  <T_{\m}^{\n}> \sim   \d_{\m}^{\n}$. However, the
identification in $\p$ breaks the underlying symmetry picking out $\p$
as a preferred direction.   $  <T_{\m}^{\n}>$ is traceless and conserved. One
should
note, however, that in analogy to the Casimir effect the energy density is
negative.

For the   rotating black hole, the stress tensor {\tmn}  becomes
\eqn\stress{\eqalign{
 <T_t^{\, t}> = & {1\over 4\pi} \sum_{n =1}^{\infty}
 \bigl ( (\cp + 2\cm -3)r^2 - 2 (\cm -1)l^2 \bigr )  { c_n \,\over |d_n|^{5/2}}
\cr
 <T_r^{\, r}> = & {1\over 4\pi} \sum_{n =1}^{\infty}
 \bigl( (\cp -\cm)r^2 + (\cm -1)l^2 \bigr ) { c_n \,\over |d_n|^{5/2}}
\cr
 <T_{\p}^{\,\p}> = & {1\over 4\pi}  \sum_{n =1}^{\infty}
\bigl ( - (2\cp + \cm -3)r^2 + (\cm -1)l^2 \bigr ) { c_n \,\over |d_n|^{5/2}}
\cr
<T_t^{\,\p} >  = & \,
 {3\over 4\pi} \sum_{n =1}^{\infty} \sinh\,n\a_+ \sinh\,n\a_-
 {{r^2 / l^2}-1\over   |d_n|^{5/2}} l \cr
  <T_t^{\, r}> = & <T_{r}^{\, \p}>  =0 \cr
   &c_n \equiv  \cp + \cm +2\cr
  & d_n    \equiv |x - \L^n x|^2 =    2 (\cosh\,n\a_+ - \cosh\,n\a_-)r^2 + 2 (
\cosh\,n\a_- -1)l^2  \cr
}}
with  $\a_{\pm}$   given in {\ab}. In the $J=0$ $(\a_- =0)$ limit,  {\stress}
reduces to
{\stressjzero}.
Recall that in   $(t, r, \p)$ coordinates, the   outer   and inner horizons are
located at $r=l$ and $r=0$.  Outside the inner horizon, where  $d_n$  is
positive and the infinite sums
converge exponentially,   $<T_{\m\n}>$ is  smooth.  The inner horizon,
in terms of the imbedding coordinates, is the surface $r^2 = T_1^2 - X_1^2 =0$
 corresponding to the lightcone in the $1+1$ space $M_1 .$ Inside the
horizon,  $ \r = r^2 = T_1^2 - X_1^2$ becomes negative,  and the
 denominators   $d_n$   in {\stress} vanish   on a sequence of timelike
surfaces
\eqn\pol{
\r   = \rho_n, \quad \rho_n\equiv -
{\cosh\,n\a_- -1\over \cosh\,n\a_+ - \cosh\,n\a_-} l^2, \quad M>J/l .  }

As we now demonstrate, the $n$th surface in {\pol} consists of points
  $x^a$ connected    to their image $\L^n x $ by a null geodesic and  is known
as a polarized hypersurface \ref\kt{
S.-W. Kim and K. S. Thorne, {\it Phys. Rev. D} {\bf 43} (1991) 3929.}.
Since $x$ and  $\L^n x $ are identified in the black hole solution,
 the connecting null geodesic  is self-intersecting.
   In {\Ads3}, geodesics are  the analogs of great circles on
 ordinary spheres. In other words, they are  curves   which also lie
on a two dimensional plane passing through the origin in the
 four dimensional imbedding space.
 Two points $x$ and $y$ are connected by a  spacelike, lightlike,  or
 timelike  geodesic depending on   whether $x^a y_a <-l^2,$ $x^a y_a =-l^2,$ or
 $-l^2< x^a y_a < l^2 $ respectively
{\ref\oneill{
B. O'Neill, {\it Semi-Riemannian Geometry}   (Academic Press, New York,
1983).}}.
 [Points with  $x^a y_a >l^2 $ lie on different branches of a hyperboloid and,
therefore, are not connected by {\it any} geodesic.]
 Since a point $x$  on the $n$th polarized hypersurface satisfies
$ d_n    = |x - \L^n x|^2 =0 $ implying $x^a (\L^n)_{ab} x^b = -l^2$,
 $x$ and $\L^n x$ are connected by a null geodesic.  As one approaches a
 polarized hypersurface {\pol}  from a geodesic distance $s$, $<T_{\m}^{\n}>$
diverges
as $s^{-5/2} .$ Since these  surfaces in the  $n\ra\infty$ limit approach
 the inner horizon, $r=0$,   the stress tensor will diverge there.
[It should be noted   that $<T_{\m\n}>$ is in fact finite at the
 inner horizon as it is approached from the outside. This is due to the fact
that though
  each of the polarized hypersurfaces contains null geodesics,  the
 inner horizon itself   does not   and is   said to be non-compactly
generated.]
  One can  estimate the back reaction due to the diverging stress tensor
by   substituting $<T_{\m\n}>$  into the field equation. Integrating twice,
 one finds that the   metric perturbation    diverges as $\delta \gmn \sim
s^{-1/2}$
on each of the polarized hypersurfaces. This suggests that the inner horizon is
 quantum mechanically unstable against formation of a curvature singularity.

 For the extremal case $(M=J/l)$,   the stress tensor {\stress}
becomes
\eqn\extremal{\eqalign{
 <T_t^{\, t}> = & K( 3r^2 - 2l^2) \cr
 <T_r^{\, r}> = &K l^2\cr
 <T_{\p}^{\, \p}> = &-K (3r^2  -l^2)\cr
<T_{t}^{\, \p} >  = & {3\over 2}K ({r^2\over l^2}-1 )l \cr
  <T_r^{\, t}> =& <T_r^{\,\p}>  =0 \cr
 K     \equiv & {\sqrt{2}\over 16\pi  l^5} \sum_{n =1}^{\infty}
 { \cosh  \,n\pi \sqrt{2M}  +1\over (  \cosh\,n\pi \sqrt{2M}  - 1)^{3/2}} . \cr
}}
 For $M=J/l >>1$, one has $K   \approx    l^{-5} e^{-\pi \sqrt{M/2}}.$
  Note that in contrast to the non-extremal case, {\extremal}
 is smooth everywhere but diverges asymptotically.

In this paper, we studied the  stress tensor for a propagating quantum field in
the $2+1$
black hole. Considering the relatively simple geometric structure of the
black hole solution, one would hope that further investigation would lead
to a greater understanding of its  quantum properties.

\bigbreak\bigskip\bigskip\centerline{\bf Acknowledgements}\nobreak
I would like to thank
Gary Gibbons, Stephen Hawking, Miguel Ortiz, and Yoav Peleg for  useful
discussions.  I am grateful to  M. Ortiz for
referring me to  {\ais}.
I also wish  to acknowledge   the financial support of the SERC.
After completion of this paper, I received  two preprints, Akita Junior
College preprint AJC-HEP-19  (August 1993)
by  K. Shiraishi and
T. Maki and M.I.T. preprint CTP-2243 (September 1993) by G. Lifschytz and M.
Ortiz,
in which the quantum stress tensor for the non-rotating
black hole is calculated.
\baselineskip=30pt
 \listrefs
\end